\documentclass[aps,twocolumn,showpacs]{revtex4}
\usepackage{graphicx}
\usepackage{amsfonts}
\usepackage{amssymb}
\usepackage{amsbsy}
\usepackage{amsmath}
\usepackage{mathrsfs}
\usepackage{latexsym}
\usepackage{natbib}
\usepackage{bm}
\usepackage{subfigure}
\usepackage{color}

\def\x{\mathbf{x}}
\def\y{\mathbf{y}}

\def\kr{\kappa}

\def\ksg{\mathrm{\varkappa}}

\def\rs{r_s}
\def\rstar{r_{\star}}
\def\scriplus{\mathscr{I}^{+}}
\def\scriminus{\mathscr{I}^{-}}

\def\observerminus{\mathbb{O}^{-}}
\def\observerplus{\mathbb{O}^{+}}

\begin{document}

\title{An exact derivation of the Hawking effect in canonical formulation}

\author{Subhajit Barman}
\email{sb12ip007@iiserkol.ac.in}

\author{Golam Mortuza Hossain}
\email{ghossain@iiserkol.ac.in}

\author{Chiranjeeb Singha}
\email{cs12ip026@iiserkol.ac.in}

\affiliation{ Department of Physical Sciences, 
Indian Institute of Science Education and Research Kolkata,
Mohanpur - 741 246, WB, India }
 

\date{\today}

\begin{abstract}

The Hawking effect is one of the most extensively studied topics in modern 
physics. Yet it remains relatively under-explored within the framework of 
canonical quantization. The key difficulty lies in the fact that the Hawking 
effect is principally understood using the relation between the ingoing modes 
which leave past null infinity and the outgoing modes which arrive at future 
null infinity. Naturally, these modes are described using advanced and retarded 
null coordinates instead of the usual Schwarzschild coordinates. However, null 
coordinates do not lead to a true Hamiltonian that describes the evolution of 
these modes. In order to overcome these hurdles in a canonical formulation, we 
introduce here a set of near-null coordinates which allows one to perform an 
exact Hamiltonian-based derivation of the Hawking effect. This derivation opens 
up an avenue to explore the Hawking effect using different canonical 
quantization methods such as polymer quantization.

\end{abstract}

\maketitle

\section{Introduction}

The \emph{Hawking effect} \cite{hawking1975} is one of the most remarkable 
results obtained by employing quantum field theory in curved spacetime 
\cite{book:Birrell,book:Parker,book:Mukhanov} where an asymptotic observer in 
future finds a thermal emission from a black hole. The thermal emissions are 
usually associated with the systems with very large number of microstates. 
However, a classical black hole which is a solution of Einstein's  general 
relativity \cite{book:Schutz,book:carroll,book:wald,Fulling:1989nb} 
can be described by only a handful of parameters. This puzzling aspect often 
leads one to believe that the study of Hawking effect might allow one to 
understand the possible microstates of a black hole which are expected to arise 
from a possible, yet unknown, quantum theory of gravity. This has led to an 
extensive set of studies on the Hawking effect in different contexts 
\cite{Lambert:2013uaa, Jacobson:2003vx, Kiefer:2002fp, Traschen:1999zr, 
DEWITT1975295, Ford:1997hb, Hollands:2014eia,Padmanabhan:2009vy, 
Chakraborty:2015nwa,Chakraborty:2017pmn,Helfer:2003va,Carlip:2014pma,
Fulling1987135,Hinton:1982, Parikh:1999mf,Visser:2001kq, 
Singleton:2011vh,Bhattacharya:2013tq, Singh:2013pxf,Lapedes:1977ip, 
Davies:1974th, Wald1975,Singh:2014paa,Dray1985,Kawai:2013mda, 
Ho:2015fja,Jacobson:2012ei,PhysRevD.46.2486,Hartle:1976tp}.

However, despite being one of the most extensively studied topics in modern 
physics, the study of Hawking effect itself remains relatively under-explored 
within the canonical quantization framework. The key reason behind the
difficulty in canonical formulation is the basic tenet through which one 
realizes the thermal emission. In particular, the thermal nature of the Hawking 
quanta is realized using the relation between the modes which leave the past 
null infinity as ingoing null rays and the modes which arrive at the future 
null infinity as outgoing null rays. As expected, instead of the regular 
Schwarzschild coordinates, the usage of the advanced and retarded \emph{null 
coordinates} then becomes quite crucial in the derivation of the Hawking effect. 
However, null coordinates do not lead to a true Hamiltonian that describes the 
evolution of these modes incongruous to our need for studying the system 
(nevertheless, see 
\cite{Torre:1985rw,Dirac:1949cp,Harindranath:1996hq,dInverno:2006wzl}).
This in turn creates hurdles for performing an extensive study of the Hawking 
effect using the canonical quantization framework.

In the context of polymer quantization of matter field, recently it has been 
argued that the Unruh effect \cite{Fulling:1972md,Unruh:1976db,Crispino:2007eb} 
may get altered significantly due to the existence of a new length scale akin to 
the Planck length  \cite{Hossain:2014fma,Hossain:2015xqa,Hossain:2016klt}. 
Polymer quantization \cite{Ashtekar:2002sn,Halvorson-2004-35} is a canonical 
quantization method which is used in loop quantum gravity 
\cite{Ashtekar:2004eh,Rovelli2004quantum,Thiemann2007modern}. Given the 
similarity of techniques employed in the study of the Unruh effect and the 
Hawking effect, naturally one then asks whether polymer quantization would also 
alter the Hawking effect.

Therefore, it has become imperative to pursue the study of the Hawking effect 
using the framework of canonical quantization. In this article, we introduce one 
such framework. In particular, we  introduce here a set of \emph{near-null 
coordinates} that allows one to closely follow the basic tenets of the Hawking 
effect and  to perform an exact Hamiltonian-based derivation of it. In an 
earlier canonical attempt using the Lema\^itre coordinates by Melnikov and 
Weinstein \cite{Melnikov:2001ex}, the Hawking effect is understood indirectly 
through the property of the Green's function rather than the expectation value 
of the associated number operator. To the best of our knowledge there doesn't 
yet exist any exact derivation of the thermal spectrum for Hawking radiation in 
canonical formulation.

In the Section \ref{Hawking-radiation-review}, we briefly review the key aspects 
of the standard derivation of the Hawking effect \cite{hawking1975}. In 
particular, a massless, free scalar field is considered for describing the 
Hawking quanta. Additionally, a collapsing shell of matter is considered whose 
eventual collapse leads to the formation of the black hole. The corresponding 
black hole spacetime is taken to be the Schwarzschild spacetime. Furthermore, 
one considers a set of two observers: one at the past null infinity and the 
other observer at the future null infinity. The observer at the past null 
infinity considers a set of ingoing modes which are specified by the 
\emph{advanced} null coordinate. On the other hand, the observer at the future 
null infinity studies the outgoing modes which are specified by the 
\emph{retarded} null coordinate. By using the relation between the advanced 
and retarded null coordinates, one computes the relevant Bogoliubov 
transformation coefficients. This in turn allows one to express the vacuum 
expectation value of the number operator associated with the Hawking quanta. The 
spectrum of these quanta turns out to be thermal in nature. The corresponding 
temperature is proportional to the \emph{surface gravity} at the Schwarzschild 
horizon and referred to as the Hawking temperature.

In the section \ref{Canonical-formulation}, we begin by describing the 
properties of the matter Hamiltonian for a massless, free scalar field in a 
general globally hyperbolic spacetime. In order to derive the Hawking effect 
within a canonical formulation, then we introduce a pair of \emph{near-null} 
coordinates. These new coordinates are used by a set of two different 
observers mainly in the asymptotic regions near past and future null infinities 
respectively. Subsequently, we derive the relation between the intervals along 
the spatial hypersurfaces which are used by these two observers.

In order to perform canonical quantization of the scalar field we consider the 
Fourier modes of the field. Later, we compute the Bogoliubov coefficients that 
relate the different Fourier modes of the two different observers. These 
coefficients are then used to compute the vacuum expectation value of the 
Hamiltonian operator for the Fourier modes as seen by the observer near future 
null infinity in the vacuum state of the observer near past null infinity. We 
identify the Hawking radiation as the characteristic outgoing radiation whose 
existences are tied with the non-zero values of the surface gravity at the event 
horizon. This leads to an exact expression for the Hawking formula and the 
corresponding Hawking temperature.

\section{Hawking Radiation}\label{Hawking-radiation-review}

In the standard derivation of the Hawking effect \cite{hawking1975} one 
considers a collapsing shell of matter whose eventual collapse leads to the 
formation of the black hole. In addition, one also considers a massless scalar 
field to describe the Hawking quanta. However, the detailed dynamics of the 
collapsing shell of matter is not important for the derivation of the Hawking 
radiation.

\subsection{Schwarzschild spacetime}

We consider the resultant spacetime, after the collapse of the matter shell to 
a black hole, to be described by the Schwarzschild geometry. In particular, the 
spacetime metric for an observer in the asymptotic future is given by
\begin{equation}\label{SchwarzschildMetric0}
ds^2 = - \Omega dt^2 + \Omega^{-1} dr^2 
+ r^2 d\theta^2 + r^2 \sin\theta^2 d\phi^2 ~,
\end{equation}
where $\Omega = \left(1- r_s /r\right)$ and $r_s = 2 G M$ is the 
Schwarzschild radius associated with the metric. Here we use 
\emph{natural units} such that $c=\hbar=1$. We note that the metric 
(\ref{SchwarzschildMetric0}) can also be used by an observer in the asymptotic 
past by taking the limit $\rs\to0$, when there was no black hole.

By defining the so-called \emph{tortoise coordinate} $\rstar$ such that $d\rstar 
= \Omega^{-1} dr$, one may reduce the Schwarzschild metric 
(\ref{SchwarzschildMetric0}) to the form
\begin{equation}\label{SchwarzschildMetric}
ds^2 = g_{\mu\nu}dx^{\mu}dx^{\nu} =  \Omega \left[ - dt^2 + d\rstar^2 \right] 
  + r^2 d\theta^2 + r^2 \sin\theta^2 d\phi^2 ~. 
\end{equation}
The choice of tortoise coordinate transforms $t-r$ plane of the
Schwarzschild geometry to become \emph{conformally flat}. By a suitable 
choice of constant of integration, $\rstar$ can be explicitly written as
\begin{equation}\label{TortoiseCoordinate}
\rstar = r + r_s \ln \left(\frac{r}{r_{s}}-1\right)  ~.
\end{equation}
For later convenience, we define the \emph{advanced} and \emph{retarded} null 
coordinates $v$ and $u$ respectively as
\begin{equation}\label{AdvancedRetardedNullCoordinates}
v = t + \rstar  ~~;~~  u = t - \rstar ~~.
\end{equation}
The Schwarzschild metric (\ref{SchwarzschildMetric}) in terms of these null 
coordinates can then be expressed as
\begin{equation}\label{SchwarzschildMetricNullCoordinates}
ds^2 =  - \Omega~ du~ dv + r^2 d\theta^2 + r^2 \sin\theta^2 
d\phi^2 ~. 
\end{equation}
%


\begin{figure}
\includegraphics[width=5cm]{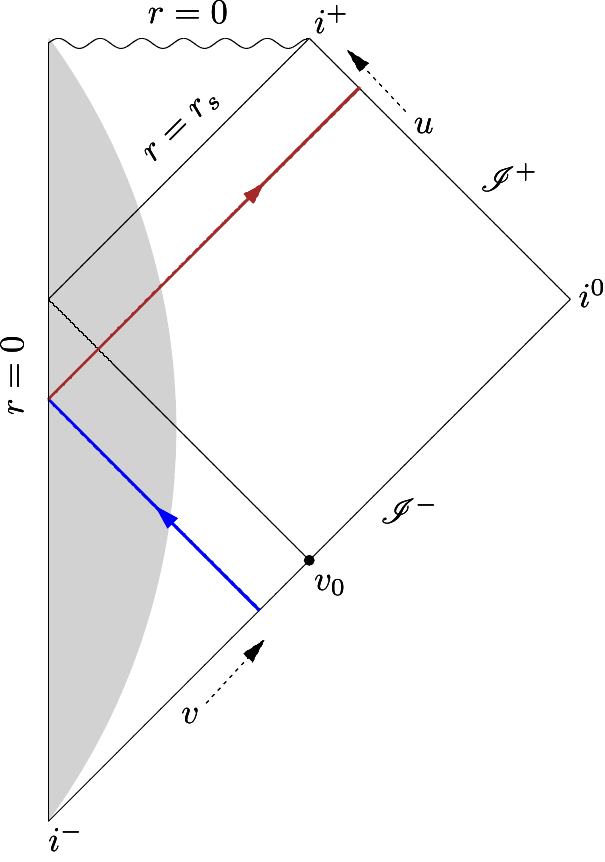}
\caption{The standard Penrose diagram which is used to describe the Hawking 
effect. The shaded region depicts the qualitative evolution of a collapsing 
shell of matter whose collapse leads to the formation of the black hole. The 
ingoing null ray departs from the past null infinity $\scriminus$ where as the 
outgoing null ray arrives at the future null infinity $\scriplus$. }
\label{fig:BHPenroseDiagram} 
\end{figure}

\subsection{Massless scalar field}

In order to describe the Hawking quanta, we consider a minimally coupled, 
massless scalar field $\Phi(x)$, described by the action
\begin{equation}\label{ScalarActionFull}
S_{\Phi} = \int d^{4}x \left[ -\frac{1}{2} \sqrt{-g} 
g^{\mu \nu} \nabla_{\mu}\Phi(x) \nabla_{\nu}\Phi(x) \right] ~.
\end{equation}
The general solutions to the Klein-Gordon field equation $\Box \Phi(x) = 0$ 
in the Schwarzschild spacetime (\ref{SchwarzschildMetric}) can be expressed as
\begin{equation}\label{ScalarFieldSolutionFull}
\Phi(x) = \sum_{\omega l m} \frac{c_{\omega lm}}{r}~ 
\tilde{f}_{\omega}(\rstar)~ e^{-i\omega(t\pm\rstar)} ~ Y_{lm}(\theta,\phi) ~,
\end{equation}
where $c_{\omega lm}$ are some constants, $\tilde{f}_{\omega}(\rstar)$ are 
solutions to the \emph{reduced} radial equation and $Y_{lm}(\theta,\phi)$ are 
the regular \emph{spherical harmonics}. In particular, at a large distance 
\emph{i.e.} for $r \gg \omega^{-1}$, the function $\tilde{f}_{\omega}(\rstar)$ 
becomes a constant.

\subsection{Creation and annihilation operators}

In order to realize the Hawking effect, it is crucial to consider two 
asymptotic observers: one is at \emph{past null infinity} $\scriminus$ and the 
other is at \emph{future null infinity} $\scriplus$ 
(see FIG. \ref{fig:BHPenroseDiagram}). In particular, with respect to the past 
observer at $\scriminus$, the scalar field operator can be expressed as 
\begin{equation}\label{ScalarFieldPastOperator}
\hat{\Phi}(x) = \sum_{\omega} \left[ 
{f}_{\omega} \hat{a}_{\omega} + {f}^{*}_{\omega} \hat{a}^{\dagger}_{\omega} 
\right] ~,
\end{equation}
where the set of \emph{ingoing} solutions to field equation $\{{f}_{\omega}\}$ 
forms a complete family on $\scriminus$ along with the inner product
$(-i/2)\int_{S}d\Sigma^{a} \left({f}_{\omega} \nabla_a 
{f}^{*}_{\omega'} -
{f}^{*}_{\omega'} \nabla_a {f}_{\omega} \right) = \delta_{\omega\omega'}$ 
where $S=\scriminus$.
Further, in order to render the corresponding inner product \emph{positive 
definite}, the only positive frequency modes $\{{f}_{\omega}\}$, with respect 
to a canonical affine parameter along $\scriminus$, are chosen. The positive 
frequency ingoing solutions near the past null infinity $\scriminus$ can be 
explicitly written as
\begin{equation}\label{IngoingSolution}
{f}_{\omega}(v) = \frac{1}{\sqrt{2\pi\omega}} ~ r^{-1}~ 
e^{-i\omega v} ~Y_{lm}(\theta,\phi) ~.
\end{equation}
The operators $\hat{a}^{\dagger}_{\omega}$ and $\hat{a}_{\omega}$ denote the 
creation and annihilation operators respectively. The corresponding vacuum state 
$|0_{-}\rangle$ is defined as
\begin{equation}\label{VacuumMinus}
\hat{a}_{\omega}~|0_{-}\rangle = 0 ~.
\end{equation}

Similarly, for an observer in asymptotic future, the scalar field operator 
can be expressed as 
\begin{equation}\label{ScalarFieldFutureOperator}
\hat{\Phi}(x) = 
\sum_{\omega} \left[ {p}_{\omega} \hat{b}_{\omega} + {p}^{*}_{\omega} 
\hat{b}^{\dagger}_{\omega} \right]
+
\sum_{\omega} \left[ {q}_{\omega} \hat{c}_{\omega} + {q}^{*}_{\omega} 
\hat{c}^{\dagger}_{\omega} \right]
~,
\end{equation}
where field solutions $\{{p}_{\omega}\}$ are purely \emph{outgoing} and 
given by
\begin{equation}\label{OutgoingSolution}
{p}_{\omega}(u) = \frac{1}{\sqrt{2\pi\omega}} ~ r^{-1}~ 
e^{-i\omega u} ~Y_{lm}(\theta,\phi) ~.
\end{equation}
These solutions (\ref{OutgoingSolution}) have zero Cauchy data on the event 
horizon. On the other hand, field solutions $\{{q}_{\omega}\}$ have zero Cauchy 
data on the future null infinity $\scriplus$. The operators 
($\hat{b}^{\dagger}_{\omega}$, $\hat{b}_{\omega}$) and 
($\hat{c}^{\dagger}_{\omega}$, $\hat{c}_{\omega}$) are the creation and 
annihilation operator pairs in the respective domain. The corresponding inner 
products are 
$(-i/2)\int_{S}d\Sigma^{a} \left({p}_{\omega} \nabla_a {p}^{*}_{\omega'} 
-{p}^{*}_{\omega'} \nabla_a {p}_{\omega} \right) = \delta_{\omega\omega'}$ 
with the integration surface being $S = \scriplus$ and 
$(-i/2)\int_{S}d\Sigma^{a} 
\left({q}_{\omega} \nabla_a {q}^{*}_{\omega'} -{q}^{*}_{\omega'} \nabla_a 
{q}_{\omega} \right) = \delta_{\omega\omega'}$ 
where $S$ is the event horizon. As earlier, the set of solutions 
$\{{p}_{\omega}\}$ are considered to contain only positive frequencies
with respect to the canonical affine parameter along the null geodesic 
generator on $\scriplus$.

\subsection{Relation between null coordinates $v$ and $u$}

An essential input that leads to the emergence of the Hawking effect is the 
relation between the null coordinates of the asymptotic observers at the past 
and future null infinities. In particular, using the relation between an affine 
parameter interval along the future null infinity $\scriplus$ and the 
corresponding interval on the past null infinity $\scriminus$, one can show that
\begin{equation}\label{Relation:v0vu0v}
(v_0 - v) \approx -2\rs~ e^{-(u_0 - u)/2 \rs} ~~,
\end{equation}
where $u^0$ and $v^0$ denote some pivotal points on $\scriplus$ and 
$\scriminus$ respectively. A way to understand the origin of the relation 
(\ref{Relation:v0vu0v}) is to use the following arguments. By 
considering a pivotal point $v^0$ on $\scriminus$, an interval along 
$\scriminus$ can be expressed as
\begin{equation}\label{ScriMinusVInterval}
(v^0 - v)_{|\scriminus} = 2(\rstar^0 -\rstar)_{|\scriminus}  ~,
\end{equation}
where $\rstar^0$ is the tortoise coordinate corresponding to the point $v^0$. 
Similarly, we can express an interval along $\scriplus$ as 
\begin{equation}\label{ScriPlusUInterval}
(u^0 - u)_{|\scriplus} = - 2(\rstar^0 -\rstar)_{|\scriplus} ~,
\end{equation}
where $u^0$ is a pivotal point on $\scriplus$ and $\rstar^0$ is the 
corresponding value of the tortoise coordinate. However, there is a key 
difference between the coordinate $\rstar$ as used by each of these two 
observers. Firstly, there was no black hole when the relevant \emph{ingoing} 
modes had departed from $\scriminus$. So for the observer at $\scriminus$ we 
must take $\rs\to0$ limit in the expression of tortoise coordinate $\rstar$. 
This in turn reduces the interval (\ref{ScriMinusVInterval}) to
\begin{equation}\label{ScriMinusVInterval2}
(v^0 - v)_{|\scriminus} = 2(r^0 -r)_{|\scriminus} \equiv \Delta ~,
\end{equation}
where $\Delta$ is taken to be a positive interval along the past null infinity 
$\scriminus$. On the other hand, when the outgoing modes arrive at the future 
null infinity $\scriplus$, the black hole horizon has already formed with
non-zero Schwarzschild radius $\rs$. Therefore, using the 
radial coordinate $r$, the interval (\ref{ScriPlusUInterval}) along $\scriplus$ 
can be expressed  as
\begin{equation}\label{ScriMinusUInterval}
(u^0 - u)_{|\scriplus} = \Delta + 
2\rs \ln \left(1 + \frac{\Delta}{\Delta_0}\right) ~,
\end{equation}
where we have defined $\Delta_0 \equiv 2(r^0 - \rs)_{|\scriplus}$ and have
identified the interval $-2(r^0 -r)_{|\scriplus}$ as $\Delta$ using 
\emph{geometric optics} approximation. By choosing the pivotal values  $v^0 = 
-\Delta_0$ and $u^0 = v^0 - 2\rs \ln (-v^0/2\rs)$ we can simplify their relation 
as 
\begin{equation}\label{Relation:AdvancedRetardedNullCoordinates}
- \frac{u}{2\rs} = - \frac{v}{2 \rs}  +  \ln \left(-\frac{v}{2 \rs}\right)  ~.
\end{equation}
With the given choices of the pivotal values, $\ln(-v/2\rs)$ term will dominate 
over $(-v/2\rs)$ term in the region where $|v| \ll 2\rs$. It turns out that the 
relevant modes for Hawking radiation are precisely those modes which originate 
from the region $|v| \ll 2\rs$ on $\scriminus$ . Therefore, in this region 
one can approximate the relation 
(\ref{Relation:AdvancedRetardedNullCoordinates}) as
\begin{equation}\label{Relation:AdvancedRetardedNullCoordinatesApprox}
v \approx -2\rs~ e^{-u/2 \rs} ~~.
\end{equation}
The relation (\ref{Relation:AdvancedRetardedNullCoordinatesApprox}) can be 
identified with the relation (\ref{Relation:v0vu0v}) with suitable choices of 
the pivotal values.  We shall use similar arguments for finding the analogous 
relation in canonical formulation.

\subsection{Bogoliubov coefficients and number operator}

Being a complete basis, one can express the outgoing modes $p_{\omega}$ in 
terms of the ingoing modes $\{{f}_{\omega}\}$ and $\{{f}^{*}_{\omega}\}$ as
\begin{equation}\label{BogoliubovTransformationHawking}
{p}_{\omega}(u) =  \sum_{\omega'} \left[ \alpha_{\omega\omega'} {f}_{\omega'}(v)
+ \beta_{\omega\omega'} {f}^{*}_{\omega'}(v) \right] ~.
\end{equation}
Due to the mixing of modes, the vacuum state $|0_{-}\rangle$ of the observer at 
the past null infinity $\scriminus$, is no longer annihilated by the 
annihilation operator $\hat{b}_{\omega}$ of the observer at future null 
infinity 
$\scriplus$ \emph{i.e.} $\hat{b}_{\omega}~|0_{-}\rangle \ne 0$. The expectation 
value of the number operator corresponding to the observer at the future null 
infinity $\scriplus$, in the vacuum state corresponding to the observer at 
past null infinity $\scriminus$, can be expressed as
\begin{equation}\label{NumberVEVDefinitionHawking}
{N}_{\omega} \equiv 
\langle 0_{-}| \hat{b}^{\dagger}_{\omega} \hat{b}_{\omega} |0_{-}\rangle
=  \sum_{\omega'} |\beta_{\omega\omega'}|^2 ~.
\end{equation}
The relation (\ref{BogoliubovTransformationHawking}) between the modes of these 
two observers and the relation 
(\ref{Relation:AdvancedRetardedNullCoordinatesApprox}) between their 
coordinates are used to explicitly evaluate the Bogoliubov transformation
coefficient $\beta_{\omega\omega'}$. This in turn leads the expectation 
value of the number operator to become
\begin{equation}\label{NumberVEVHawking}
{N}_{\omega} = \frac{1}{e^{2\pi\omega/\ksg} - 1}  ~,
\end{equation}
where $\ksg=1/(2\rs)$ is the \emph{surface gravity} at the horizon. The 
equation 
(\ref{NumberVEVHawking}) corresponds to the spectrum of blackbody radiation for 
bosons at the temperature $T_H = \ksg/(2\pi k_B) = 1/(8\pi G M k_B)$. This 
phenomena of blackbody radiation perceived by the observer at the future null 
infinity $\scriplus$ in a black hole spacetime is referred to as the Hawking 
effect. The corresponding temperature $T_H$ is called the Hawking temperature.

\section{Canonical formulation}\label{Canonical-formulation}

A key structural step that leads to the derivation of the Hawking effect in the 
covariant formulation, is the Bogoliubov transformation between the solutions of 
the field which are functions of advanced null coordinate $v$ (\emph{i.e.} 
ingoing modes) and the functions of retarded null coordinate $u$ (\emph{i.e.} 
outgoing modes). Furthermore, in order to evaluate these transformation 
coefficients explicitly it is essential to have the relation between the null 
coordinates $v$ and $u$ along the past null infinity $\scriminus$ and the future 
null infinity $\scriplus$ respectively. However, despite being intuitively 
appealing, these coordinates, being null, pose challenges in the canonical 
formulation. In particular, null coordinates do not lead to a true Hamiltonian 
that describes the evolution of these modes. Therefore, one must look for other 
suitable coordinates to perform a Hamiltonian-based derivation of the Hawking 
effect.

\subsection{General Scalar Field Hamiltonian}

In this subsection, we briefly review few key aspects of a 3+1 
spatio-temporal decomposition \cite{book:3plus1} of a general globally 
hyperbolic spacetime. After the decomposition, the invariant distance element 
can be expressed as
\begin{equation}\label{GeneralMetricIn31}
ds^2 = -N^2 dt^2 + q_{ab}(N^a dt+d\x^{a})(N^b dt + d\x^{b}) ~,
\end{equation}
where $q_{ab}$ denotes the \emph{spatial metric}, $N$ is the \emph{lapse 
function} and $N^a$ is the \emph{shift vector} associated with the foliation of 
the spacetime into the spatial hypersurfaces $\Sigma_t$, labeled by the 
time parameter $t$.
The full scalar field Hamiltonian corresponding to the action 
(\ref{ScalarActionFull}) can be expressed as
\begin{equation}\label{GeneralScalarFieldHamiltonianFull}
H_{\Phi} = \int d^{3}x \left[ N \mathcal{H} + N^a \mathcal{D}_a \right] ~,
\end{equation}
where the scalar Hamiltonian density $\mathcal{H}$ and the diffeomorphism 
generator $\mathcal{D}_a$ are given by
\begin{equation}\label{ScalarHamiltonianAndDiffeomorphismFull}
\mathcal{H} = \frac{\Pi^2}{2\sqrt{q}} + 
\frac{\sqrt{q}}{2} q^{ab} \partial_{a}\Phi \partial_{b}\Phi  
~~;~~  
\mathcal{D}_a = \Pi~ \partial_a \Phi
.
\end{equation}
The determinant of the spatial metric $q_{ab}$ is denoted by $q$. The Poisson 
brackets between the field $\Phi$ and its conjugate   momentum $\Pi$ can be 
written as
\begin{equation}\label{PoissonBracketFull}
\{\Phi(t,\x), \Pi(t,\y)\} = \delta^3(\x-\y) ~.
\end{equation}
Using the Hamilton's equation of motion, it is straightforward to check that 
the field momentum $\Pi$ can be expressed as 
\begin{equation}\label{FieldMomentumGeneral}
\Pi = \frac{\sqrt{q}}{N} 
\left[\partial_{t}\Phi - N^a \partial_{a}\Phi \right] ~.
\end{equation}

\subsection{1+1 dimensional reduced action}

We have already noted that the Hawking effect is essentially connected with 
the structure of the Schwarzschild metric in the $t-r$ plane. Therefore, in 
order to  simplify the analysis, here we reduce the $3+1$ dimensional scalar 
field action (\ref{ScalarActionFull}) to an $1+1$ dimensional action by 
integrating out the angular coordinates $\theta$ and $\phi$. In particular, 
by considering the form of the general field solution 
(\ref{ScalarFieldSolutionFull}), we make an ansatz for the scalar field of the 
form
\begin{equation}\label{ScalarFieldYLMSubstitution}
\Phi(x^j,\theta,\phi) = \sum_{lm}  \tilde{\varphi}_{lm}(x^j) 
Y_{lm}(\theta,\phi) ~,
\end{equation}
where $x^j = t,\rstar$. By substituting this general ansatz 
(\ref{ScalarFieldYLMSubstitution})  of $\Phi(x^j,\theta,\phi)$ into the action 
(\ref{ScalarActionFull}) and integrating over the angular coordinates, we get 
the reduced action of the form
\begin{eqnarray}
S_{\Phi} &=& \sum_{l,m} \int dt d\rstar \left[ 
\frac{1}{2}(\partial_{t}\tilde{\varphi}_{lm})^{2}-\frac{1}{2}(\partial_{\rstar}
\tilde{\varphi}_{lm})^{2} \right. \nonumber\\
      &-& \left. \frac{\Omega}{2r^{2}}  \{\Omega+l(l+1)\} 
(\tilde{\varphi}_{lm})^{2} +\frac{\Omega}{r}
\tilde{\varphi}_{lm} \partial_{\rstar} \tilde{\varphi}_{lm}  \right].
\label{ReducedScalarActionFull}
\end{eqnarray}
The Hawking effect is understood using the relation between the modes of scalar 
fields as seen by the observers at the past and future null infinities. 
Therefore, with respect to these two observers, one can simplify the reduced 
action (\ref{ReducedScalarActionFull}) by dropping the terms which explicitly 
contain inverse powers of $r$ and are comparatively smaller at large radial 
distances (\emph{i.e.} $r\to\infty$). The remaining terms in the simplified 
action then become independent of $l$ and $m$. One may redefine the reduced 
scalar field $\varphi \propto \tilde{\varphi}_{lm}$, such that the simplified 
action can be viewed as a scalar field action in an $(1+1)$ dimensional 
Schwarzschild spacetime, given by
\begin{equation}\label{ReducedScalarAction2D}
S_{\varphi} = \int d^{2}x\;\left[ -\frac{1}{2} \sqrt{-\bar{g}} \bar{g}^{ij} 
\partial_{i}\varphi \partial_{j} \varphi \right] ~,
\end{equation}
where $\bar{g}_{ij}$ is the corresponding $(1+1)$ dimensional metric. We shall 
use this reduced action (\ref{ReducedScalarAction2D}) for further computations.

\subsection{Near-null coordinates}

In the canonical formulation, the field dynamics can be viewed as the `time 
evolution' the field modes on the `spatial hypersurfaces'. Clearly, the advanced 
and retarded null coordinates are not suitable in the canonical formulation and 
one must look for coordinates which are not null. Firstly, we note that the 
\emph{ingoing} field solutions (\ref{IngoingSolution}) have a phase factor of 
the form $e^{-i\omega v}$. Along a given ingoing null trajectory the advanced 
null coordinate $v$ is constant. However, one may use the retarded null 
coordinate $u$ to parameterize the propagation along the trajectory. In other 
words, ingoing field solutions ${f}_{\omega}(v)$, using the relation 
$v=u+2\rstar$, can be viewed as if ${f}_{\omega}(u) = e^{-i\omega u } 
~{f}_{\omega}(0)$ where $u$ changes along the trajectory. Remarkably, this form 
can be compared with the time evolution of a Schrodinger wave function 
$\psi_{\omega}(\tau) = e^{-i\omega\tau}\psi_{\omega}(0)$ corresponding to a 
mechanical system with energy $\omega$ and the time coordinate $\tau$. 
Furthermore, we know that a massless, free scalar field can be mapped into a set 
of quantum mechanical harmonic oscillators by using Fourier transformation. 
Therefore, these insights suggest that in order to realize the Hawking effect in 
a canonical formulation and yet to mimic the methods of Bogoliubov 
transformations between the null coordinates, we could define a timelike 
coordinate by slightly deforming the retarded null coordinate $u$ and define a  
spacelike coordinate by slightly deforming the advanced null coordinate $v$ for 
an appropriate observer near the past null infinity $\scriminus$. Similar 
arguments can be made for an observer near the future null infinity $\scriplus$ 
using the \emph{outgoing} field solutions.

Following above insights, we define a new set of coordinates to be used by an 
observer near the past null infinity $\scriminus$ as
\begin{equation}\label{NearNullCoordinatesMinus}
\tau_{-} = t - (1-\epsilon)\rstar ~~;~~ \xi_{-} = -t - (1+\epsilon)\rstar  ~,
\end{equation}
where $\epsilon$ is a real-valued parameter such that $\tau_{-}$ and $\xi_{-}$ 
are timelike and spacelike coordinates respectively. Here we choose the 
parameter $\epsilon$ to be a small and positive such that $\epsilon \gg 
\epsilon^2$. This choice of parameter mimics the basic tenets of the Hawking 
effect very closely. However, in principle, one can choose the values of the 
parameter in the domain $0<\epsilon<2$ which preserves the timelike 
characteristic of the coordinate $\tau_{-}$. In any case, the final result will 
be independent of the explicit values of $\epsilon$. We refer this observer as 
$\observerminus$.

\begin{figure}
\includegraphics[width=9cm]{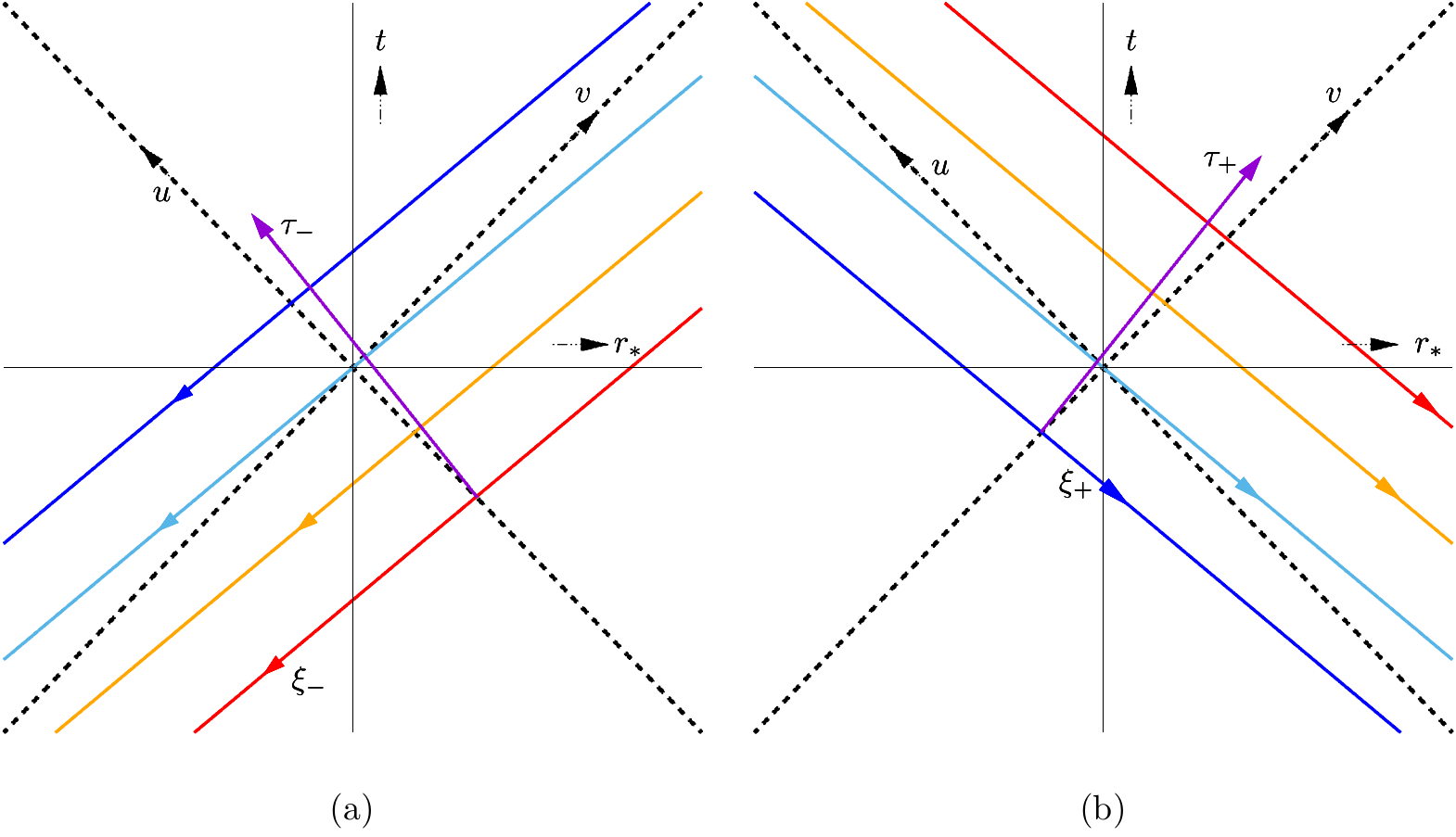}
\caption{(a) The near-null coordinates for the observer $\observerminus$.
(b) The near-null coordinates for the observer $\observerplus$.}
\label{fig:NearNull} 
\end{figure}

Similarly for an observer near the future null infinity $\scriplus$, we define 
another set of timelike and spacelike coordinates $\tau_{+}$ and $\xi_{+}$ 
as
\begin{equation}\label{NearNullCoordinatesPlus}
\tau_{+} = t + (1-\epsilon)\rstar ~~;~~ \xi_{+} = -t + (1+\epsilon)\rstar  ~.
\end{equation}
As earlier, we refer this observer as $\observerplus$. We may note from the 
equation (\ref{NearNullCoordinatesMinus}) and the equation 
(\ref{NearNullCoordinatesPlus}) that one can algebraically transform these two 
set of the coordinates from one to another by simply substituting 
$\rstar \to -\rstar$. Further, for later convenience, we have chosen the 
directions of $\xi_{-}$ and $\xi_{+}$ to be the opposite of the directions of 
$v$ and $u$ coordinates respectively when $\epsilon=0$ is set (See FIG. 
\ref{fig:NearNull}).


\subsection{Relation between spatial coordinates $\xi_{-}$ and $\xi_{+}$}

In order to explicitly realize the Hawking effect in covariant formulation, one 
needs to compute Bogoliubov transformation coefficient between the ingoing 
field solutions (\emph{i.e.} functions of $v$) and outgoing field solutions 
(\emph{i.e.} functions of $u$). The key input that is required to evaluate 
these coefficients is the relation 
(\ref{Relation:AdvancedRetardedNullCoordinatesApprox}) between the null
coordinates of two different observers. We show here that one can derive an
analogous relation between the spatial coordinates $\xi_{-}$ and $\xi_{+}$. As 
earlier, let us consider a pivotal point $\xi_{-}^0$ on a $\tau_{-} = 
constant$ surface as seen by the observer $\observerminus$ near the past 
null infinity. Then an interval along this surface can be expressed as
\begin{equation}\label{ScriMinusXiMinuxInterval}
(\xi_{-} - \xi_{-}^0)_{|\tau_{-}} = 2(\rstar^0-\rstar)_{|\tau_{-}} =
2(r^0-r)_{|\tau_{-}} \equiv \Delta ~, 
\end{equation}
where $r^0$ corresponds to the pivotal value $\xi_{-}^0$ and the interval 
$\Delta$ is positive. In the equation (\ref{ScriMinusXiMinuxInterval})
we have used the fact that for the observer near the past null infinity there 
was no black hole \emph{i.e.} $\rs\to0$. For the observer $\observerplus$ we can 
express an interval on a $\tau_{+} = constant$ surface near future null infinity 
as
\begin{equation}\label{ScriMinusXiPlusInterval}
(\xi_{+} - \xi_{+}^0)_{|\tau_{+}} 
= \Delta + 2\rs \ln \left(1 + \frac{\Delta}{\Delta_0}\right)
 ~,
\end{equation}
where $\Delta_0 \equiv 2(r_0 - \rs)_{|\tau_{+}}$ and using geometric optics 
approximation we have identified the interval $2(r - r^0)_{|\tau_{+}}$ also as 
$\Delta$.  By choosing the pivotal values 
$\xi_{+}^0 = \xi_{-}^0 + 2\rs \ln (\Delta_0/2\rs)$ and $\xi_{-}^0 = \Delta_0$, 
we can simplify the relation  between the spatial coordinates $\xi_{-}$ and 
$\xi_{+}$ as
\begin{equation}\label{Relation:ximinusxiplusExact}
\xi_{+} = \xi_{-} + 2\rs \ln \left(\frac{\xi_{-}}{2\rs}\right) ~~.
\end{equation}
The modes which are seen as Hawking radiation to the observer $\observerplus$, 
propagate out from a region where $|\xi_{-}| \ll 2\rs$ on a constant $\tau_{-}$ 
surface as seen by the observer $\observerminus$ near the past null infinity. 
For these modes, the equation (\ref{Relation:ximinusxiplusExact}) can be 
approximated as
\begin{equation}\label{Relation:ximinusxiplus}
\xi_{-} \approx  2 \rs~e^{\xi_{+}/2 \rs} ~~.
\end{equation}
The equation (\ref{Relation:ximinusxiplus}) plays the similar role as the 
relation (\ref{Relation:AdvancedRetardedNullCoordinatesApprox}) between the 
advanced and retarded null coordinates.


\begin{figure}
\includegraphics[width=5cm]{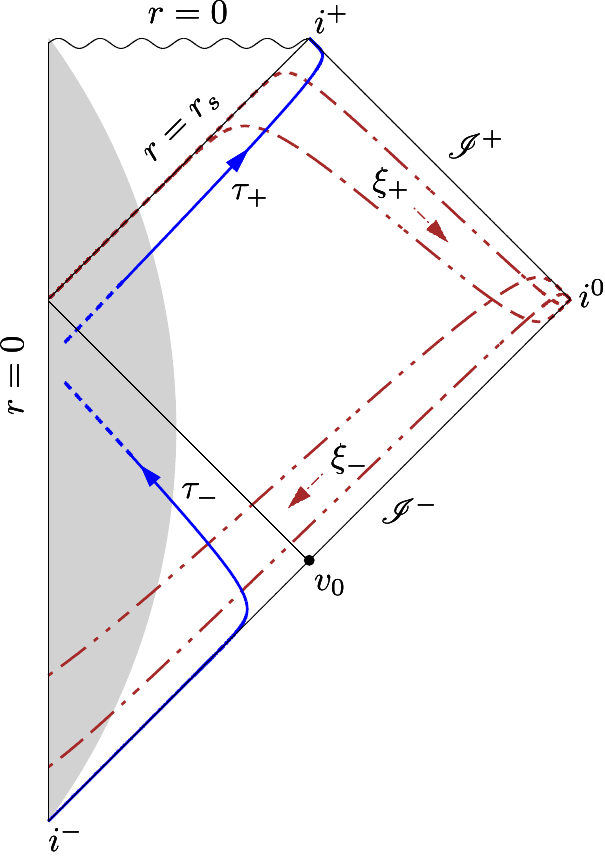}
\caption{The spatial hypersurfaces and the temporal directions in near-null 
coordinates drawn on a Penrose diagram together with a collapsing shell of 
matter. 
}
\label{fig:NearNullMinusPenrose} 
\end{figure}

\subsection{Field Hamiltonian for Observer $\observerminus$}

We have seen that the Hawking effect is crucially connected with 
the structure of the Schwarzschild metric in the $t-r$ plane. This 
1+1 dimensional metric when viewed from the coordinate system used by
the observer $\observerminus$, appears to be of the form
\begin{equation}\label{NearNullMetricMinus}
ds^2 = g^{-}_{\mu\nu}dx^{\mu}dx^{\nu} = \frac{\epsilon\; \Omega}{2}
\left[ -  d\tau_{-}^2 + d\xi_{-}^2 
+\frac{2}{\epsilon} d\tau_{-} d\xi_{-} \right]  ~. 
\end{equation}
In the spacetime geometry (\ref{NearNullMetricMinus}), the reduced action 
(\ref{ReducedScalarAction2D}) for a minimally coupled massless scalar field  can 
also be expressed as
\begin{equation}\label{ReducedScalarAction2DFlat}
S_{\varphi} =  \int d\tau_{-}  d\xi_{-} \left[-\frac{1}{2} \sqrt{-g^{0}} 
g^{0\mu\nu} \partial_{\mu}\varphi \partial_{\nu} \varphi \right]  ~,
\end{equation}
where $g^{-}_{\mu\nu} = (\epsilon ~\Omega/2) g^{0}_{\mu\nu}$. The metric 
$g^{0}_{\mu\nu}$ although has off-diagonal terms but being flat it allows 
one to use the machinery of the Fock quantization. 

Following the general form (\ref{GeneralScalarFieldHamiltonianFull}),
the scalar field Hamiltonian in 1+1 dimensions, as seen by the observer 
$\observerminus$, can be expressed as
\begin{equation}\label{ScalarHamiltonianFullMinus}
H_{\varphi} = \int d\xi_{-}  \frac{1}{\epsilon}  \left[
\left\{ \frac{\Pi^2}{2}  + \frac{1}{2}  (\partial_{\xi_{-}}\varphi)^2 \right\} 
+ \Pi~ \partial_{\xi_{-}} \varphi \right] ~,
\end{equation}
where the \emph{lapse function} $N = 1/\epsilon$, the \emph{shift vector} $N^1 = 
1/\epsilon$ and the determinant of the spatial metric $q = 1$. The Poisson 
bracket between the field $\varphi$ and its conjugate momentum $\Pi$ can be 
written as
\begin{equation}\label{PoissonBracketMinus}
\{\varphi(\tau_{-},\xi_{-}), \Pi(\tau_{-},\xi_{-}')\} 
= \delta(\xi_{-} - \xi_{-}') ~.
\end{equation}
Using the equations of motion, the field momentum $\Pi$ can be expressed 
as 
\begin{equation}\label{FieldMomentumMinus}
\Pi(\tau_{-},\xi_{-}) = \epsilon (\partial_{\tau_{-}}\varphi) - 
(\partial_{\xi_{-}}\varphi) ~.
\end{equation}

\subsection{Field Hamiltonian for Observer $\observerplus$}

With respect to the observer $\observerplus$ the metric for the 1+1 dimensional 
spacetime is given by
\begin{equation}\label{NearNullMetricPlus}
ds^2 = g^{+}_{\mu\nu}dx^{\mu}dx^{\nu} = \frac{\epsilon\; \Omega}{2}
\left[ - d\tau_{+}^2 + d\xi_{+}^2 
+ \frac{2}{\epsilon} d\tau_{+} d\xi_{+} \right]  ~. 
\end{equation}
As earlier, by performing a conformal transformation of the metric as 
$g^{+}_{\mu\nu} = (\epsilon ~\Omega/2) g^{0}_{\mu\nu}$, one can express the 
scalar field Hamiltonian for the observer $\observerplus$ as
\begin{equation}\label{ScalarHamiltonianFullPlus}
H_{\varphi} = \int d\xi_{+}  \frac{1}{\epsilon} \left[
\left\{ \frac{\Pi^2}{2}  + \frac{1}{2}  (\partial_{\xi_{+}}\varphi)^2 \right\} 
+ \Pi~ \partial_{\xi_{+}} \varphi \right] ~,
\end{equation}
where the \emph{lapse function} $N = 1/\epsilon$, the \emph{shift 
vector} $N^1 = 1/\epsilon$ and the determinant of the spatial metric
$q = 1$.  The corresponding Poisson bracket is given by
\begin{equation}\label{PoissonBracketPlus}
\{\varphi(\tau_{+},\xi_{+}), \Pi(\tau_{+},\xi_{+}')\} = \delta(\xi_{+} - 
\xi_{+}') ~.
\end{equation}
Similar to the expression (\ref{FieldMomentumMinus}), field momentum for the
observer $\observerplus$ can be expressed as 
\begin{equation}\label{FieldMomentumPlus}
\Pi(\tau_{+},\xi_{+}) = \epsilon (\partial_{\tau_{+}}\varphi) 
- (\partial_{\xi_{+}}\varphi) ~.
\end{equation}

\subsection{Fourier modes for Observer $\observerminus$ }

In order to perform the canonical quantization of the scalar field we follow 
the method as used in \cite{Hossain:2010eb}. Firstly, we define the Fourier 
modes for the scalar field with respect to the observer $\observerminus$, as
\begin{equation}\label{FourierModesDefinitionMinus}
\varphi =  \frac{1}{\sqrt{V_{-}}}\sum_{k} \tilde{\phi}_{k} e^{i k \xi_{-}} ~;~
\Pi =  \frac{1}{\sqrt{V_{-}}} \sum_{k} \sqrt{q}~ \tilde{\pi}_{k} 
e^{i k \xi_{-}} ~,
\end{equation}
where $\tilde{\phi}_{k} = \tilde{\phi}_{k} (\tau_{-})$, 
$\tilde{\pi}_{k} = \tilde{\pi}_{k} (\tau_{-})$ are the complex-valued mode 
functions. The spatial volume $V_{-} = \int d\xi_{-}\sqrt{q}$ is formally 
divergent. Therefore, to avoid dealing with explicitly divergent quantity, we 
choose a fiducial box with finite volume as 
\begin{equation}\label{SpatialVoumeMinus}
V_{-} = \int_{\xi_{-}^L}^{\xi_{-}^R} d\xi_{-}\sqrt{q} = {\xi_{-}^R} - 
{\xi_{-}^L} \equiv L_{-} ~.
\end{equation}
We shall see later that with the given definition of the Fourier modes 
(\ref{FourierModesDefinitionMinus}), the fiducial volume $V_{-}$ will drop out 
from the expression of the Hamiltonian of the modes. Given the finiteness of 
the 
fiducial volume, the Kronecker delta can be expressed as
\begin{equation}\label{KroneckerDeltasMinus}
\int d\xi_{-}\sqrt{q}  e^{i(k-k')\xi_{-}} = V_{-} \delta_{k,k'} ~,
\end{equation}
whereas the Dirac delta can be written as
\begin{equation}\label{DiracDeltasMinus}
\sum_k e^{ik(\xi_{-}-\xi_{-}')} = V_{-} \delta(\xi_{-}-\xi_{-}')/\sqrt{q} ~.
\end{equation}
The equation (\ref{KroneckerDeltasMinus}) and the equation 
(\ref{DiracDeltasMinus}) together imply that wave-vector $k \in \{k_r\}$ 
where $k_r = 2\pi r/L_{-}$ with $r$ being a non-zero integer.

The scalar field Hamiltonian (\ref{ScalarHamiltonianFullMinus}) for the observer
$\observerminus$ can be expressed in terms of the Fourier modes as 
$H_{\varphi} = \sum_k (\mathcal{H}_k^{-} + \mathcal{D}_k^{-})/\epsilon$ where
Hamiltonian density $\mathcal{H}_k^{-}$ for the $k^{th}$ mode is
\begin{equation}\label{FourierHamiltonianMinus}
\mathcal{H}_k^{-} = \frac{1}{2} \tilde{\pi}_{k}  \tilde{\pi}_{-k}
+ \frac{1}{2} |k|^2 \tilde{\phi}_{k}  \tilde{\phi}_{-k} ~,
\end{equation}
and diffeomorphism generator $\mathcal{D}_k^{-}$ is
\begin{equation}\label{FourierDiffeomorphismMinus}
\mathcal{D}_k^{-} = - \frac{i k}{2} \left( 
\tilde{\pi}_{k} \tilde{\phi}_{-k} - 
\tilde{\pi}_{-k} \tilde{\phi}_{k} \right) ~.
\end{equation}
The Poisson bracket between the Fourier modes of the scalar field and its 
conjugate momentum is given by
\begin{equation}\label{FourierPoissonBracketMinus}
\{\tilde{\phi}_{k}, \tilde{\pi}_{-k'}\} = \delta_{k,k'} ~.
\end{equation}

\subsection{Fourier modes for Observer $\observerplus$ }

In parallel to the case of the observer $\observerminus$, we define the 
Fourier modes for the scalar field as seen by the observer $\observerplus$
as
\begin{equation}\label{FourierModesDefinitionPlus}
\varphi =  \frac{1}{\sqrt{V_{+}}}\sum_{\kr} \tilde{\phi}_{\kr} 
e^{i \kr \xi_{+}} ~;~
\Pi =  \frac{1}{\sqrt{V_{+}}} \sum_{\kr} \sqrt{q}~ \tilde{\pi}_{\kr} 
e^{i \kr \xi_{+}} ~,
\end{equation}
where $\tilde{\phi}_{\kr} = \tilde{\phi}_{\kr} (\tau_{+})$, 
$\tilde{\pi}_{\kr} = \tilde{\pi}_{\kr} (\tau_{+})$ are the complex-valued mode 
functions. The spatial volume $V_{+}$ is also chosen to be that of a fiducial 
box with finite volume, given by
\begin{equation}\label{SpatialVoumePlus}
V_{+} = \int_{\xi_{+}^L}^{\xi_{+}^R} d\xi_{+}\sqrt{q} = {\xi_{+}^R} - 
{\xi_{+}^L} \equiv L_{+} ~.
\end{equation}
Using the appropriate representations of the Kronecker delta and the Dirac 
delta, as in the equations (\ref{KroneckerDeltasMinus}) and 
(\ref{DiracDeltasMinus}), the scalar field Hamiltonian 
(\ref{ScalarHamiltonianFullPlus}) can be expressed in terms of the Fourier 
modes 
as $H_{\varphi} = \sum_{\kr} (\mathcal{H}_{\kr}^{+} + 
\mathcal{D}_{\kr}^{+})/\epsilon$ where the Hamiltonian density is
\begin{equation}\label{FourierHamiltonianPlus}
\mathcal{H}_{\kr}^{+} = \frac{1}{2} \tilde{\pi}_{\kr}  \tilde{\pi}_{-\kr}
+ \frac{1}{2} |\kr|^2 \tilde{\phi}_{\kr}  \tilde{\phi}_{-\kr} ~,
\end{equation}
the diffeomorphism generator is
\begin{equation}\label{FourierDiffeomorphismPlus}
\mathcal{D}_{\kr}^{+} =  
 -\frac{i \kr}{2} \left( \tilde{\pi}_{\kr} \tilde{\phi}_{-\kr} -
 \tilde{\pi}_{-\kr} \tilde{\phi}_{\kr} \right)  ~,
\end{equation}
and the corresponding Poisson bracket is given by
\begin{equation}\label{FourierPoissonBracketPlus}
\{\tilde{\phi}_{\kr}, \tilde{\pi}_{-\kr'}\} = \delta_{\kr,\kr'} ~.
\end{equation}

\subsection{Relation between Fourier modes}

In order to establish the relation between the Hamiltonian densities of Fourier 
modes for these two observers, we need to find the relation between the 
individual modes of the field and their conjugate momenta. Being a scalar, the 
matter field can be expressed as $\varphi(\tau_{-},\xi_{-}) = 
\varphi(\tau_{+},\xi_{+})$ where the coordinates can be viewed as $\tau_{-} = 
\tau_{-}(\tau_{+},\xi_{+})$ and $\xi_{-} = \xi_{-}(\tau_{+},\xi_{+})$ in 
general. However, the relation between the field momentum 
(\ref{FieldMomentumMinus}) and (\ref{FieldMomentumPlus}) are slightly involved. 
Firstly, we note that the observer $\observerminus$ deals with the field modes 
which are \emph{ingoing} modes. For these modes $v = t + \rstar = 
(\epsilon/2)\tau_{-} - (1-\epsilon/2) \xi_{-}$  is \emph{constant}. Similarly, 
the relevant modes for the observer $\observerplus$ are the \emph{outgoing} 
modes. For these modes, $u = t - \rstar = (\epsilon/2)\tau_{+} - (1-\epsilon/2) 
\xi_{+}$ is \emph{constant}. Therefore, the temporal derivative term in the 
expression of the field momentum is not independent and can be related to the 
spatial derivative term. This property holds true for both the observers. This 
in turn leads to a relation between the field momenta as $\Pi(\tau_{+},\xi_{+}) 
= (\partial \xi_{-}/\partial \xi_{+}) \Pi(\tau_{-},\xi_{-})$.
We use these relations between the field and the conjugate momentum, as seen by 
two different observers, to establish the relation between the Fourier modes. In 
particular, by choosing spatial hypersurfaces for the observers $\observerminus$ 
and $\observerplus$, labeled by fixed $\tau_{-}^0$ and $\tau_{+}^0$ 
respectively, we can relate the Fourier modes of the field as
\begin{equation}\label{FieldModesRelation}
\tilde{\phi}_{\kr} = \sum_{k} \tilde{\phi}_{k} F_{0}(k,-\kr) ~,
\end{equation}
where $\tilde{\phi}_{\kr} = \tilde{\phi}_{\kr}(\tau_{+}^0)$ and 
$\tilde{\phi}_{k} = \tilde{\phi}_{k}(\tau_{-}^0)$. Similarly, the Fourier modes 
for field momenta can also be related as
\begin{equation}\label{FieldMomentaModesRelation}
\tilde{\pi}_{\kr} =  \sum_{k} \tilde{\pi}_{k} F_{1}(k,-\kr)  ~,
\end{equation}
where $\tilde{\pi}_{\kr} = \tilde{\pi}_{\kr}(\tau_{+}^0)$ and 
$\tilde{\pi}_{k} = \tilde{\pi}_{k}(\tau_{-}^0)$. The coefficient function
$F_{m}(k,\kr)$ is given by
\begin{equation}\label{FFunctionGeneral}
F_{m}(k,\kr) = \frac{1}{\sqrt{V_{-} V_{+}}} 
\int d\xi_{+} ~e^{m\xi_{+}/2 \rs} ~e^{i k \xi_{-} + 
i \kr \xi_{+}} ~,
\end{equation}
where $m=0,1$. The coefficient functions $F_{m}(k,\kr)$ play the similar role 
as the Bogoliubov coefficients in the covariant formulation.

\subsection{Regularization of Bogoliubov coefficients}

Like the standard Bogoliubov coefficients, these coefficient functions are 
also formally divergent and one needs to employ some regularization techniques 
to render them finite. Here we follow the regularization techniques which is 
similar to the one used in \cite{Hossain:2014fma}. Firstly, one introduces 
non-oscillatory regulator with a small parameter $\delta$ in the expression 
(\ref{FFunctionGeneral}), as follows
\begin{equation}\label{FFunctionRegulated}
F_{m}^{\delta}(k,\kr) = \frac{1}{\sqrt{V_{-} V_{+}}} 
\int d\xi_{+} \left[\frac{e^{(m+\delta)\xi_{+}/2 \rs}}{d_m} \right]
e^{i k \xi_{-} + i \kr \xi_{+}}  ~,
\end{equation}
where $d_m = (1 - i\delta m/2\rs\kr)$. One can check that regulated expression 
(\ref{FFunctionRegulated}) reduces to the exact expression 
(\ref{FFunctionGeneral}) when the parameter $\delta$ is removed
\emph{i.e.} $\lim_{\delta\to0} F_{m}^{\delta}(k,\kr) = F_{m}(k,\kr)$.

In order to evaluate the integral (\ref{FFunctionRegulated}), one performs a
change of variable as $z \equiv |k| \xi_{-}$. This in turn leads to 
\begin{equation}\label{FFunctionEvaluated}
F_{m}^{\delta}(\pm|k|,\kr) = \frac{(2\rs)^{-\beta-m} |k|^{-\beta-m-1} }
{d_m ~\sqrt{V_{-} V_{+}}} I_{\pm}(\beta+m) ~,
\end{equation}
where $\beta = (2i\kr\rs + \delta - 1)$. The integral $I_{\pm}(\beta) = \int dz 
~e^{\pm i z} z^{\beta}$ can be explicitly expressed in terms Gamma function 
$\Gamma(\beta+1)$ by analytic continuation either in the upper or lower half of 
the complex plane, depending on the sign of $k$. The evaluated expression is 
given by
\begin{equation}\label{IntegralIPM}
I_{\pm}(\beta) = e^{\pm i \pi(\beta+1)/2} ~\Gamma(\beta+1)  ~.
\end{equation}
In order to express the integral (\ref{IntegralIPM}) in terms of \emph{complete}
Gamma function, one needs to add two boundary terms 
$\Delta I^{L} = \int_0^{|k|\xi_{-}^L} dz ~e^{\pm i z} z^{\beta}$ and 
$\Delta I^{R} = \int_{|k|\xi_{-}^R}^{\infty} dz ~e^{\pm i z} z^{\beta}$. Both 
of these terms vanish identically when one removes the volume regulator by 
taking the limit $\xi_{-}^L\to0$ and $\xi_{-}^R\to\infty$. 
For later convenience, we write down two key relations between the different 
forms of $F_{m}^{\delta}(\pm|k|,\kr)$, for different values of the parameter, as
\begin{equation}\label{F0F0Relation}
F_{0}^{\delta}(-|k|,\kr) = e^{2\pi\rs\kr - i\delta\pi}~F_{0}^{\delta}(|k|,\kr) 
~,
\end{equation}
and 
\begin{equation}\label{F0F1Relation}
F_{1}^{\delta}(\pm|k|,\kr) = \mp \frac{\kr}{|k|}~F_{0}^{\delta}(\pm|k|,\kr) ~.
\end{equation}

\subsection{Consistency relation between the regulators}          

We have introduced two sets of regulator so far. One set is to regulate the 
volumes of the flat spatial slices through the means of finite ($\xi_{-}^L$, 
$\xi_{-}^R$) and ($\xi_{+}^L$, $\xi_{+}^R$). The relation 
(\ref{Relation:ximinusxiplus}) implies that they are related among 
themselves. Secondly, we have introduced the parameter $\delta$ as the integral 
regulator. However, it turns out that these two sets of regulators can not be 
chosen independently in order to ensure the consistency of the Poisson brackets 
for both the observers.

In particular, the requirement that both Poisson brackets 
$\{\tilde{\phi}_{k},\tilde{\pi}_{-k'}\} = \delta_{k,k'}$ and
$\{\tilde{\phi}_{\kr},\tilde{\pi}_{-\kr'}\} = \delta_{\kr,\kr'}$
are simultaneously satisfied, leads to a consistency condition on the
coefficient functions as
\begin{equation}\label{PoissonBracketConsistencyCondition}
\sum_{k>0} \left[ F_{0}(k,-\kr)F_{1}(-k,\kr) 
+ F_{0}(-k,-\kr)F_{1}(k,\kr) \right] = 1 ~.
\end{equation}
In terms of the regulated expression (\ref{FFunctionRegulated}) of 
$F_{m}(k,\kr)$, the  equation (\ref{PoissonBracketConsistencyCondition}) demands
\begin{equation}\label{PoissonBracketConsistencyCondition2}
\frac{(\kr\rs/\pi) |\Gamma(2i\kr\rs + \delta)|^2}
{(e^{2\pi\kr\rs}-e^{-2\pi\kr\rs})^{-1}}
= \frac{(V_{+}/2\rs) (4\pi\rs/V_{-})^{2\delta} }{\zeta(1+2\delta)} ~,
\end{equation}
where the \emph{Riemann zeta function} $\zeta(1+2\delta) = \sum_{r=1}^{\infty} 
r^{-1(1+2\delta)}$. Using the Gamma function identity 
$\Gamma(z)\Gamma(1-z) = \pi/\sin\pi z$, the zeta function identity 
$\lim_{s\to0}[s~ \zeta(1+s)] = 1$ and the equation 
(\ref{Relation:ximinusxiplus}), it is straightforward to show that 
the volume regulator $\xi_{-}^L$ and integral regulator $\delta$ should be
varied together as $\xi_{-}^L/2\rs \simeq 2\pi e^{-1/2\delta}$. In other words,
the consistency condition (\ref{PoissonBracketConsistencyCondition}) implies
that these two regulators are not independent of each other.

\subsection{Relation between Hamiltonian densities and Diffeomorphism 
generators}

Using the relations (\ref{FieldModesRelation}, \ref{FieldMomentaModesRelation}) 
between the Fourier modes of the field and their conjugate momenta, we can 
establish the relation between the corresponding Hamiltonian densities and the 
diffeomorphism generators. Furthermore, by using the relations  
(\ref{F0F0Relation}) and (\ref{F0F1Relation}), we can express the Hamiltonian 
density $\mathcal{H}_{\kr}^{+}$ as
\begin{equation}\label{ModesHamiltonianRelations0}
\mathcal{H}_{\kr}^{+} = h_{\kr}^1 + (e^{4\pi\rs\kr} + 1)  
\sum_{k>0} \left(\frac{\kr}{k}\right)^2 ~|F_{0}(k,\kr)|^2 ~\mathcal{H}_k^{-}  ~,
\end{equation}
where 
$
h_{\kr}^1 =  \sum_{k\neq k'} [ 
\frac{1}{2} F_{1}(k,-\kr) F_{1}(-k',\kr) ~ \tilde{\pi}_{k}  \tilde{\pi}_{-k'} 
+ \frac{1}{2} |\kr|^2 F_{0}(k,-\kr) F_{0}(-k',\kr) ~ 
\tilde{\phi}_{k} \tilde{\phi}_{-k'} ] ~.
$
Similarly, we can express the diffeomorphism generators of the Fourier modes  
corresponding to the observer $\observerplus$, as
\begin{equation}\label{ModesDiffeomorphismRelations0}
\mathcal{D}_{\kr}^{+} = d_{\kr}^1 + (e^{4\pi\rs\kr} + 1)  
\sum_{k>0} \left(\frac{\kr}{k}\right)^2 ~|F_{0}(k,\kr)|^2 ~\mathcal{D}_k^{-}  ~,
\end{equation}
where 
$
d_{\kr}^1 =  \sum_{k\neq k'} (-i\kr/2) [  F_{1}(k,-\kr) F_{0}(-k',\kr) ~ 
\tilde{\pi}_{k}  \tilde{\phi}_{-k'} 
- F_{1}(-k,\kr) F_{0}(k',-\kr) ~ \tilde{\pi}_{-k} \tilde{\phi}_{k'} ]~.
$
We note that both the terms $h_{\kr}^1$ and $d_{\kr}^1$ are linear in 
Fourier modes or their conjugate momenta. Therefore, the vacuum expectation 
values of the corresponding operators in the quantum theory will vanish 
for these terms.

\subsection{Reality condition on diffeomorphism generators}

In order to represent the Hawking quanta, we have considered here a 
real-valued scalar field \emph{i.e.} $\varphi^{*}(x) = \varphi(x)$. This 
property in turn imposes a \emph{reality condition} on the complex-valued 
Fourier modes as $\tilde{\phi}^{*}_{k} = \tilde{\phi}_{-k}$. In general, we can 
express a complex-valued mode function as $\tilde{\phi}_{k} = \phi_{k}^r + 
i~\phi_{k}^i$ where  $\phi_{k}^r$ and $\phi_{k}^i$ both are real-valued 
functions. Similarly, we can express the Fourier modes of the conjugate field 
momentum as $\tilde{\pi}_{k} = \pi_{k}^r + i~\pi_{k}^i$ which are also 
subjected to the reality condition $\tilde{\pi}^{*}_{k} = \tilde{\pi}_{-k}$. 
Therefore, unless one imposes this reality condition appropriately, there would 
be double counting of the degrees of freedom in terms of the real-valued mode 
functions.

In order to remove this double counting and also to express the total 
Hamiltonian $(\mathcal{H}_k^{-} + \mathcal{D}_k^{-})$  in terms of the 
real-valued mode functions, here we make a \emph{choice} by setting $\phi_{k}^i 
= 0$ and $\pi_{k}^i = 0$. The key advantages of this choice are that it 
brings $\mathcal{H}_k^{-}$ (\ref{FourierHamiltonianMinus}) to the form of a 
standard Hamiltonian of a simple harmonic oscillator with real-valued 
coordinate and it also makes the diffeomorphism generator term 
$\mathcal{D}_k^{-}$ (\ref{FourierDiffeomorphismMinus}) to vanish identically. 
Therefore, by redefining the modes as $\phi_{k} \equiv \phi_{k}^r$ and  
$\pi_{-k} \equiv \pi_{k}^r$, we can reduce the Hamiltonian density and
the Poisson bracket for the observer $\observerminus$ as 
\begin{equation}\label{FourierHamiltonianMinusReal}
\mathcal{H}_k^{-} = \frac{1}{2} \pi_{k}^2 + \frac{1}{2} |k|^2\phi_{k}^2
~~;~~  \{\phi_{k}, \pi_{k'}\} = \delta_{k,k'} ~.
\end{equation}
We shall use the simplified form (\ref{FourierHamiltonianMinusReal}) 
of the Hamiltonian density for performing Fock quantization. We should mention 
here that one could also arrive at the similar conclusion by considering a 
general complex-valued scalar field ab initio \cite{book:ADas,book:Peskin}
and imposing reality condition only at the end.

\subsection{Number operator using Hamiltonian density}

The Fock quantization of the scalar field is achieved by essentially quantizing 
the real-valued Fourier modes $\phi_{k}$ by using the method of Schrodinger 
quantization. We have already seen that a massless, free scalar field can be 
viewed as a system of infinitely many decoupled harmonic oscillators. Therefore, 
the Fock space is essentially a direct product space of infinitely many quantum 
harmonic oscillators. If we denote the vacuum state for the $k^{th}$ mode as 
$|0_k\rangle$ then the Fock vacuum state for the observer $\observerminus$ can 
be expressed as $|0_{-}\rangle = \prod_{k} \otimes |0_k\rangle$. Furthermore, 
the energy spectrum of the $k^{th}$ oscillator can be written as 
$\hat{\mathcal{H}}_{k}^{-} |n_k\rangle = (n+\frac{1}{2})|k| |n_k\rangle$ where 
$n$ denotes the energy levels.

We may recall that in covariant formulation the Hawking radiation is realized 
by computing the vacuum expectation value of the number operator corresponding 
to an observer at future null infinity $\scriplus$ whereas the vacuum state is 
taken to be that of the observer at the past null infinity $\scriminus$. 
Therefore, in order to realize the Hawking effect, here we consider the vacuum 
state to be $|0_{-}\rangle$ which is the vacuum state of the observer 
$\observerminus$. On the other hand, we consider the matter field operators 
that correspond to the observer $\observerplus$. For such a combination, the 
vacuum expectation value of the Hamiltonian density operator $\langle  
\hat{\mathcal{H}}_{\kr}^{+} \rangle \equiv \langle 0_{-}| 
\hat{\mathcal{H}}_{\kr}^{+} |0_{-}\rangle$ of a \emph{positive} frequency mode 
\emph{i.e.} $\kr>0$, can be expressed as
\begin{equation}\label{HamiltonianPlusVEV}
\frac{\langle\hat{\mathcal{H}}_{\kr}^{+}\rangle}{\kr} =
\frac{e^{2\pi\kr/\ksg} + 1}{e^{2\pi\kr/\ksg} - 1} 
\left[ \frac{1}{\zeta(1+2\delta)}
\sum_{r=1}^{\infty} \frac{1}{r^{1+2\delta}} ~
\frac{\langle\hat{\mathcal{H}}_{k_r}^{-}\rangle}{k_r}
\right]
~,
\end{equation}
where we have used the properties of the vacuum state such that $\langle 
0_k|\hat{\phi}_{k} |0_k\rangle = 0$ and $\langle 0_k|\hat{\pi}_{k} |0_k\rangle = 
0$. 

In the Fock quantization, usually one defines the number operator by defining 
the \emph{creation} and \emph{annihilation} operators of the respective modes.
However, we show here that one can also extract the vacuum expectation 
value of the number operator directly from the vacuum expectation value of the 
Hamiltonian density operator corresponding to the given mode. This will also be 
useful for the situation where the notion of creation and annihilation operators 
may not be readily available. So we define the number density operator which 
represents the Hawking quanta as
\begin{equation}\label{NumberOperatorDefinition}
\hat{N}_{\kr} = \left[ 
\hat{\mathcal{H}}_{\kr}^{+} - \lim_{\ksg\to0} \hat{\mathcal{H}}_{\kr}^{+} 
\right] |\kr|^{-1}  ~.
\end{equation}
The definition of number operator (\ref{NumberOperatorDefinition}) makes it 
clear that the existences of these Hawking quanta are related to the non-zero 
values of the surface gravity $\ksg$ at the horizon, as we have subtracted out 
the contribution to the Hamiltonian density due to the zero surface gravity.

In Fock quantization $\langle\hat{\mathcal{H}}_{k_r}^{-}\rangle = 
\frac{1}{2}|k_r|$ for all Fourier modes. This in turns leads the vacuum 
expectation value of the number density operator 
(\ref{NumberOperatorDefinition}) to become
\begin{equation}\label{NumberOperatorVEV}
N_{\omega} = \langle \hat{N}_{\kr=\omega}\rangle 
= \frac{1}{e^{2\pi\omega/\ksg} - 1} 
= \frac{1}{e^{(4\pi\rs)\omega} - 1} ~.
\end{equation}
The expression (\ref{NumberOperatorVEV}) precisely corresponds to a blackbody 
radiation at the temperature $T_H = \ksg/(2\pi k_B) = 1/(4\pi\rs k_B)$. The 
temperature $T_H$ is known as the Hawking temperature. Clearly, it demonstrates 
that one could obtain the exact thermal spectrum for Hawking radiation using a 
Hamiltonian approach.

\section{Discussion}\label{discussion}

In summary, we have presented an exact derivation of the Hawking effect using 
the canonical formulation. In the standard covariant derivation of the Hawking 
effect, the thermal nature of the Hawking radiation is realized using the key 
relation between the modes which leave the past null infinity as ingoing null 
rays and the modes which arrive at the future null infinity as outgoing null 
rays. This key relation in essence underlies the basic tenets of the Hawking 
effect. Naturally, to describe the Hawking effect in covariant formulation, it 
is quite convenient to  use the advanced and retarded null coordinates. However, 
these null coordinates are not quite useful in the canonical formulation. In 
particular, null coordinates do not lead to a true Hamiltonian that describes 
the evolution of these modes. This in turn creates hurdles for a 
Hamiltonian-based derivation of the Hawking effect in a canonical 
quantization framework. Here we overcome these hurdles by introducing a pair of 
near-null coordinates. These new coordinates which are used by two different 
observers, are obtained by a slight deformation of the advanced and the retarded 
null coordinates.  Being structurally close to the null coordinates, these new 
coordinates allow one to follow the basic tenets of the Hawking effect very 
closely. Therefore, the presented framework in this article opens up a rather 
new avenue to explore the Hawking effect using various canonical quantization 
methods such as the polymer quantization \cite{Barman:2017a2}.

\begin{acknowledgments}
We would like to thank Gopal Sardar for many useful discussions.  SB and CS 
would like to thank IISER Kolkata for supporting this work through doctoral 
fellowships. 
\end{acknowledgments}

\end{document}